\documentclass{emulateapj}

\slugcomment{Submitted to ApJ}
\shorttitle{Accretion Onto the Supermassive Black Hole in 0957+561}
\shortauthors{R. Gil-Merino et al.}

\begin{document}

\title{Accretion Onto the Supermassive Black Hole in the High--redshift Radio--loud AGN 0957+561}

\author{Rodrigo Gil-Merino, Luis J. Goicoechea, Vyacheslav N. Shalyapin\altaffilmark{1} and 
Vittorio F. Braga}
\affil{Departamento de F\'\i sica Moderna, Universidad de Cantabria, Avda. de Los Castros s/n, 
39005 Santander, Spain}
\email{r.gilmerino@gmail.com, luis.goicoechea@unican.es}
\altaffiltext{1}{permanent address: Institute for Radiophysics and Electronics, National 
Academy of Sciences of Ukraine, 12 Proskura St., 61085 Kharkov, Ukraine}

\begin{abstract}
We present the results of our X--ray, UV and optical monitoring campaign of the first 
gravitationally lensed AGN from late 2009 to mid 2010. The trailing (B) image of the AGN 
0957+561 shows the intrinsic continuum variations that were predicted in advance based on 
observations of the leading (A) image in the $gr$ optical bands. This multiwavelength 
variability of the B image allows us to carry out a reverberation mapping analysis in the 
radio--loud AGN 0957+561 at redshift $z$ = 1.41. We find that the $U$--band and $r$--band light 
curves are highly correlated with the $g$--band record, leading and trailing it by 3 $\pm$ 1 
days ($U$ band) and 4 $\pm$ 1 days ($r$ band). These 1$\sigma$ measurements are consistent with 
a scenario in which flares originated in the immediate vicinity of the supermassive black hole 
are thermally reprocessed in a standard accretion disk at $\sim$ 10--20 Schwarzschild radii from 
the central dark object. We also report that the light curve for the X--ray emission with 
power--law spectrum is delayed with respect to those in the $Ugr$ bands by $\sim$ 32 days. Hence, 
the central driving source can not be a standard corona emitting the observed power--law X--rays. 
This result is also supported by X--ray reprocessing simulations and the absence of X--ray 
reflection features in the spectrum of 0957+561. We plausibly interpret the lack of reflection 
and the 32--day delay as evidence for a power--law X--ray source in the base of the jet at a 
typical height of $\sim$ 200 Schwarzschild radii. A central EUV source would drive the 
variability of 0957+561. 
\end{abstract}


\section{Introduction}         
In the accretion paradigm for AGNs, viscous dissipation in a standard (geometrically thin and 
optically thick) accretion disk is responsible for UV--optical continuum emission 
\citep{shakura73}. Moreover, in this paradigm, the X--ray emission with power--law spectrum is 
produced in a standard corona on the rotation axis just above the black hole. The coronal X--ray 
photons are then partially reflected and thermally reprocessed into UV--optical continuum 
radiation in the disk \citep{collin91,george91}. Thus, coronal power--law X--ray fluctuations 
with timescales $\leq$ 100 days in the source rest--frame are expected to drive UV--optical 
short--timescale events \citep[e.g., see the formalism of][]{kazanas01}. However, the precise 
origin and geometrical distribution of the X--ray, UV and optical energy production are still 
largely unknown. 

Direct spatial resolution of the accretion flow in AGNs is not currently possible 
\citep[e.g.,][]{krolik99}, so we must use indirect techniques to resolve the emitting regions. 
Fortunately, reverberation (or echo) mapping is a promising time--domain technique to probe the 
accretion physics for AGNs. This is based on the analysis of time--delayed responses of 
different emitting regions to original fluctuations in a driving source 
\citep{blandford82,netzer97}. Concurrent X--ray--UV--optical continuum monitoring campaigns of 
low--redshift AGNs led to puzzling findings. In general, the fastest (days) X--ray and 
UV--optical variations matched up poorly \citep[e.g.,][]{maoz00,gaskell06,breedt09}. On the 
other hand, the slowest (tens of days) multiwavelength variability yielded ambiguous results, 
sometimes supporting the accretion paradigm \citep[e.g.,][]{breedt09}, often suggesting 
otherwise \citep[e.g.,][]{maoz00,gaskell06,arevalo09}. The situation is even less clear for AGNs 
at redshift $z \geq$ 1, since there are no X--ray--UV--optical reverberation studies of these 
objects. 

The amount of radio emission and the presence of relativistic jets divide active galactic nuclei 
in radio--quiet AGNs (RQAGNs) and radio--loud AGNs (RLAGNs). RLAGNs represent only the 10\% of 
all known AGNs. They display relativistic jets, likely launched from regions on the rotation 
axes of their central supermassive black hole \citep[e.g.,][]{urry03}. Thus, while RLAGNs having 
one jet closely aligned with the line of sight (the so--called blazars or jet--dominated RLAGNs) 
are used to probe the jet physics \citep[e.g.,][]{urry98}, RLAGNs with a relatively large 
inclination angle (accretion--dominated RLAGNs) are key tools to unveil the accretion onto black 
holes in AGNs displaying jets \citep{landt08}. 

The RLAGN \object{0957+561} at $z$ = 1.41 suffers a strong gravitational lens effect 
\citep{walsh79}. The gravitational field of a foreground galaxy cluster produces two images, A 
and B, of the same background AGN. A and B arrive at the observer at different times, with 
intrinsic flux variations in B lagging those in A by about 14 months \citep[e.g.,][]{kundic97}. 
This is an exceptional fact that offers two fundamental advantages with respect non--lensed 
AGNs: first, if we detect large flux variations in A a multiwavelength campaign can be planned 
to follow the variability in B one year later; second, if flux variations in A are indeed 
intrinsic to the AGN, similar fluctuations must also appear in B and we know that these are 
exclusively associated to the AGN physics. Moreover, radio observations of the jet in 
\object{0957+561} suggested a small beaming factor \citep{campbell94}, and hence, an 
accretion--dominated X--ray--UV--optical emission towards the observer \citep{landt08}. 

We are monitoring \object{0957+561} in the optical range since the year 2005 with the Liverpool 
Robotic Telescope (LRT), as part of a large gravitational lenses project \citep{shalyapin08}. 
Observations in late 2008 and the first semester of 2009 showed significant variations in image 
A, whose $g$--band flux increased $\sim$ 30\% just after a deep minimum. Assuming an intrinsic 
origin for the optical fluctuations in A, we were then able to predict a strong variability 
during the first semester of 2010 in image B \citep{goico09}. To take this opportunity, we 
organized an ambitious multiwavelength follow--up campaign. This included observations with the 
space--based telescopes Chandra and Swift/UVOT for X--rays and UV, respectively, and with the 
ground--based LRT in the $g$ and $r$ optical bands\footnote{We also observed 0957+561B with the 
LRT in the $i$ and $z$ bands. However, in this paper we only discuss the best data from each 
telescope. The photometric accuracy and time coverage in the $iz$ Sloan passbands are worse than 
those in the $gr$ Sloan passbands.}. In Section 2 we present these observations and the 
corresponding light curves. Preliminary records of B indicated the intrinsic origin of 
the observed variations \citep[see Fig. 2 of][]{goico11}, so the reverberation mapping analysis 
in Section 3 unveils the nature of the accretion flow and its jet connection in a distant RLAGN 
for the first time. In Section 4 we present our main conclusions.

\section{Observations and light curves}
At optical wavelengths we used the LRT. This is a ground--based fully--robotic 2m telescope 
\citep{steele04} at La Palma, Canary Islands, which is one of the world's best astronomical 
sites \citep{munoz97}. All LRT optical frames were obtained between 2009 December 26
and 2010 June 25 with the RATCam CCD camera in the $gr$ Sloan passbands (see 
Table~\ref{tbl1}). In order to get a typical signal--to--noise ratio of $\sim$ 100 for both AGN 
images for each observing night, we set the exposure times to 120 s per night in both bands. The 
pre--processing steps included in the LRT pipeline are: bias subtraction, overscan trimming and 
flatfielding. In addition, we interpolate over bad pixels using the bad pixel mask and clean 
some cosmic rays. The pre--processed frames are then analyzed using our photometric pipelines 
\citep{shalyapin08}: the crowded--field photometry pipeline produces instrumental point spread 
function (PSF) magnitudes of the AGN images, while an additional pipeline transforms 
instrumental magnitudes into Sloan Digital Sky Survey (SDSS) magnitudes. Ultimately, we turn 
magnitudes into fluxes (in mJy) using SDSS conversion 
equations\footnote{\url{http://www.sdss.org/dr7/algorithms/fluxcal.html}.}. 

We obtain 55 $g$--SDSS fluxes for each image with accuracies of 1.2\% (A) and 1.3\% (B), as 
well as 58 $r$--SDSS fluxes for each image with accuracies of 1.0\% (A) and 1.1\% (B). This 
variability database is available in tabular format at the GLENDAMA Web 
site\footnote{\url{http://grupos.unican.es/glendama/LQLMII\_DR.htm}.}. The optical light curves 
of 0957+561B are depicted in the top panel of Fig. 1 (squares and circles). As a result of the 
cosmic expansion, observed wavelengths are longer than emission wavelengths at the AGN. Thus, 
our optical observations correspond to far and middle UV continuum sources at $z$ = 1.41. 

We also used the UV/Optical Telescope (UVOT) on board the satellite Swift \citep{roming05} to 
observe \object{0957+561} in the near UV. A total of 35 $U$--band CCD frames were obtained 
between 2010 January 12 and 2010 June 1 (see Table~\ref{tbl1}), and we show one of them in Fig. 
2. The exposure 
time varied in the range of 155--1092 s, with a median value of $\sim$ 500 s. We reduce the 
observations from standard HEASoft packages\footnote{\url{http://heasarc.nasa.gov/lheasoft/}.}. 
The UVOTSOURCE task is used for performing aperture photometry on point--like objects. This 
task incorporates a coincidence--loss correction, which is important even for relatively faint 
objects \citep{li06}. Taking into account the size of the PSF (full--width at half maximum of 
about $2\farcs 5$) and the $\sim 6\arcsec$ separation of the AGN images, we choose an aperture 
radius of $3\arcsec$. However, the UVOT calibration is based on counts measured within a 
$5\arcsec$ aperture, so an aperture correction is required \citep{poole08}. This correction 
depends on the PSF of the object of interest which in turn depends on the brightness, position 
and time \citep{li06,poole08}. We estimate the aperture correction for each exposure, using the 
PSF of the nearest star to the lensed AGN. This H star has a brightness similar to that of the 
AGN images, A and B. We also check the stability of three reference bright stars close to H 
(see A, B, H and stars XFG in Fig. 2). The $U$--UVOT fluxes (in mJy) of \object{0957+561} are 
available at the GLENDAMA Web site$^3$, and the light curve of B is shown in the top panel of 
Fig. 1 (triangles). 

Observations in X--rays were obtained via Director's Discretionary Time with the space--based 
Chandra Observatory \citep{weisskopf02}. They were carried out every other week from January 17 
to June 23 in 2010 with the ACIS--S3 detector, resulting in 12 datasets evenly spaced with an 
exposure time of $\sim$ 3 ks each (see Table~\ref{tbl1}). X--ray spectra are then extracted for 
images A and B and the corresponding background fields using standard routines from the CIAO 4.3 
software\footnote{\url{http://cxc.harvard.edu/ciao/}.}. The regions for AGN spectra extraction 
are $\sim 2\farcs 5$ in radius centered at the peak of the emission; the backgrounds are $\sim 
3\arcsec$ in radius, located close to the source extraction regions. We test different 
positions for the background regions, finding no significant differences. The regions are 
always located at the same coordinates from dataset to dataset. At this stage, for extraction 
purposes, we consider the energy range from 0.1 to 11 keV in the observer rest--frame. We 
also check for possible pile--up$^5$ but we find none, surely due to the short exposure time 
of individual snapshots. 

To investigate the fiducial model underlying the AGN X--ray emission, for each AGN image, we 
combine all the individual spectra to obtain a single high signal--to--noise one. In doing 
this, we identify the features that are common to both AGN images in all frames and thus the 
emission model components that explain the X--ray sources; then we use that model to fit all 
the individual frames, studying in this way how the different components of the model vary with 
time. In the energy range 0.1--10 keV in the observer's frame, an absorbed power--law model 
\citep{chartas02} produces uncomfortable fits to the global spectra of A and B (dashed lines in 
Fig. 3). To improve the fits, we introduce a power--law plus blackbody radiation at the redshift 
of the distant AGN, both affected by Galactic neutral absorption \citep[$N_{\rm H} = 8.2 \times 
10^{19}$ cm$^{-2}$;][]{dickey90}. We then obtain 1$\sigma$ parameter values: $\Gamma$ = 1.775 
$\pm$ 0.035 and $kT$ = 0.07 $\pm$ 0.02 keV (best fit: $\chi^2$/dof = 1.03, dof = degrees of 
freedom) for A, and $\Gamma$ = 1.790 $\pm$ 0.035 and $kT$ = 0.085 $\pm$ 0.030 keV (best fit: 
$\chi^2$/dof = 0.93) for B, where $\Gamma$ is the photon index of the power--law and $kT$ is the 
blackbody temperature. There is no evidence for absorption exceeding the Galactic value, and the 
blackbody radiation is necessary to account for the soft excess detected below 0.5 keV. Both fits 
agree with each other, so we take the average values $\Gamma$ = 1.78 and $kT$ = 0.08 keV as the 
best model parameters for representing the AGN X--ray emission (solid lines in Fig. 3). We also 
check for other two--component models consisting of a power--law plus additional emission. For 
example, a two power--law model leads to worse fits for reasonable photon indices. 

The X-ray variability of the B image is obtained by fitting its individual spectra to our best 
X--ray emission model, and leaving as free parameters the normalization of the two model 
components, i.e., power--law and blackbody. We compute then the flux contribution of the 
power--law and blackbody emissions separately. We use the energy ranges 2--10 keV and $\leq$ 1 
keV for the power--law and blackbody emissions, respectively, both at the source rest-frame. 
Several tests with different two--component models give rise to light curves similar in shape to 
those from our best model. The power--law and blackbody fluxes of \object{0957+561} are available 
at the Web site of the GLENDAMA project$^3$. In the bottom panel of Fig. 1, we display two X--ray 
light curves of 0957+561B. While circles represent the observed power--law fluxes (2--10 keV 
emission), the dashed line describes the unabsorbed power--law fluxes in the same energy range. 
In Fig. 4 we also see noisy blackbody fluxes of 0957+561B. These fluxes ($\leq$ 1 keV emission) 
are substantially smaller than the power--law ones, and have large uncertainties of 30--70\%.

\section{Continuum reverberation mapping in 0957+561}

\subsection{Cross--correlation analysis}
The cross--correlation function permits a fair estimation of the time delay or time lag between 
two signals. For determining the delay between two discrete data trains, \citet{edelson88} 
introduced the discrete cross--correlation function ($DCF$), which we use in our analysis. The 
$DCF$ is characterized by a bin semisize $\alpha$, and one must check how different $\alpha$ 
values affect its form. Consecutive independent bins are separated by two times $\alpha$. 
However, there is an ambiguity in the choice of central lags of bins, so these lags can be 
shifted in an amount $< 2\alpha$ to the left or right of any initial values. The same is true 
for the discrete autocorrelation function ($DAF$). Because of such ambiguity, both the $DCF$ and 
$DAF$ are evaluated in almost continuous sets of bins. Although nearby bins are not independent,
this procedure exclusively relies on measured fluxes and correlations. Other methods such as 
interpolating (or fitting) the light curves or the $DCF$/$DAF$ yield unmeasured fluxes or 
correlations. 

The time delay is given by the centroid of the $DCF$ peak. Besides the delay, the two 
signals may differ in shape. If they are related through simple effective parameters: a flux 
offset and a multiplicative factor, the broad and irregular delay--peak should be closely traced 
by the symmetrical peak around zero lag of the $DAF$. Hence, the time--shifted peak of the $DAF$ 
can be matched to the $DCF$ peak to estimate the delay in a self–-consistent way. The time delay 
corresponds to the minimum of the so--called $\delta^2$ function, i.e. the minimum of the square 
difference between the $DCF$ and the time--shifted $DAF$ \citep[e.g.,][]{serra99}. We follow a 
standard Monte Carlo approach to generate synthetic light curves and determine time delay errors. 
We make 1000 repetitions of the experiment by adding random quantities to the original fluxes in 
the light curves. The random quantities are realisations of normal distributions around zero, 
with standard deviations equal to the errors of the fluxes. For each pair of curves in Fig. 1: 
$ab$, where $a,b = g, r, U, X$, a negative delay $\Delta \tau_{ab} = \tau_b - \tau_a <$ 0 means 
that $b$ is leading, while a positive delay $\Delta \tau_{ab} >$ 0 means the opposite ($b$ 
trailing). 

We initially focus on the $gU$ and $gr$ comparisons, i.e., the $g$--band light curve is 
compared to the other two UV--optical brightness records. Firstly, in order to measure the $gU$ 
time delay, we consider $\alpha$ = 5--25 days. We find that both the $DAF$ and $DCF$ peaks 
are smoother for longer bins, but the relationship between the two remains basically unchanged 
and their maxima always exceed 0.8. We also find that all $\delta^2$ functions indicate a 
negative delay ($U$ leading). These $\delta^2$ functions are particularly narrow for $\alpha$ = 
10--12 days (see Fig. 5), so we compute the delay and its errors using $\alpha$ = 10 days. 
Secondly, we compare the records in the $g$ and $r$ bands. The $DAF$ and $DCF$ peaks for 
$\alpha$ = 5--25 days again show maxima exceeding 0.8. However, this time the $\delta^2$
functions are consistent with a positive delay ($r$ trailing). The bin semisizes $\alpha$ = 
18--20 days produce narrow $\delta^2$ functions (see Fig. 6), and we estimate the $gr$ delay 
using $\alpha$ = 20 days. We note that a relatively long bin is required to tighten the 
$\delta^2$ curve. This could be due to the presence of a few small gaps ($\sim$ 10--20 days) in 
both light curves. 

At the redshift of \object{0957+561}, the \hbox{C\,{\sc iv}} ($\lambda$1549), 
\hbox{C\,{\sc iii]}} ($\lambda$1909) and \hbox{Mg\,{\sc ii}} ($\lambda$2798) emission lines lie 
in spectral regions covered by the $U$, $g$ and $r$ filters, respectively \citep{walsh79}. 
Therefore, one might think that the observed delays between the UV--optical light curves are 
associated with variations of lines rather than changes in the continuum. To check the 
contribution of the continuum and emission lines to our fluxes, we reduce and analyze new 
Nordic Optical Telescope (NOT) spectra of 0957+561B. These were taken on 2010 March 28, in the 
middle of our monitoring campaign (see Table~\ref{tbl1}) and covering the $gr$ filters. From the 
NOT data, we infer continuum--to--total ratios of $\sim$ 0.97 in the $g$ and $r$ bands. Thus, the 
emission line contamination is tipically about 3\%, representing a very small fraction of the 
optical fluxes.

We also study the time delays between the power--law X--ray curve and the three UV--optical 
records. The pair of curves $XU$ leads to $DCF$ peaks with maxima of $\sim$ 0.8 for $\alpha$ = 
10--25 days. However, the $\delta^2$ function is relatively broad for these bins. $DCF$ maxima 
are only reduced to $\sim$ 0.7 for longer bins with $\alpha \sim$ 30 days, but the 
corresponding $\delta^2$ curves are clearly narrower. Thus, we take $\alpha$ = 30 days to 
measure the $XU$ delay (see top panels of Fig. 7). There is an evident time shift of $\sim$ 30 
days between the $DAF$ and $DCF$ peaks, which is detected for any value of $\alpha$. This time 
shift produces a minimum in $\delta^2$ at about $-$30 days, suggesting a negative delay ($U$ 
leading) of about one month. The other two comparisons yield similar results using $\alpha$ = 
45 days (see middle and bottom panels of Fig. 7). For the two pairs of curves $Xg$ and $Xr$, 
$DCF$ maxima have values of $\sim$ 0.5--0.6 over a wide range of $\alpha$, i.e., $\alpha$ = 
10--50 days, and $\delta^2$ functions are broader than those for the $XU$ pair. 

In addition to the X--ray/UV--optical delays, we assess the significance of the relatively low
cross--correlation peaks in Fig. 7. For the $XU$ comparison, the maximum correlation is 0.7. This 
corresponds to two time series consisting of $N_X$ = 12 and $N_U$ = 35 points. Assuming a number 
of pairs of data $N = (N_X + N_U)/2 \sim$ 23, the correlation is significant at about the 99.98\% 
confidence level. In other words, uncorrelated data are very unlikely ($\sim$ 0.02\%) to produce 
a correlation $\geq$ 0.7 \citep[e.g.,][]{taylor97}. Moreover, completely uncorrelated time series 
would generate a $DCF$ noise of $\sim$ 0.08--0.09 \citep{edelson88}, which is well below the 
filled circles in the top left panel of Fig. 7. For the $Xg$ and $Xr$ comparisons, the maximum 
correlation slightly exceeds 0.5. If we consider $N \sim$ 34 ($N_g$ = 55 and $N_r$ = 58), then 
the correlation is significant at about the 99.7\% confidence level. The $DCF$ noise is $\sim$ 
0.06--0.07, and the signal--to--noise ratio in both $DCF$ peaks reaches values in the interval 
3--10.  

How do \object{0957+561} cross--correlation results fit into the accretion paradigm for AGNs and 
the associated rapid variability?. We produce 1000 synthetic light curves for each observed 
record in Fig. 1. In each synthetic curve, the observed fluxes are modified by random Gaussian 
deviations that are consistent with the measured uncertainties (see above). Thus, we obtain 1000 
delay values for each pair $gU$, $gr$, $XU$, $Xg$ and $Xr$. The corresponding delay histograms 
are drawn in Fig. 8. For the pairs $gU$ (prominent peak around $-$3 days) and $gr$ (prominent 
peak around 4 days), their delay distributions are narrower than those involving the X--ray 
record. From these narrow distributions, we infer 1$\sigma$ uncertainties (68\% confidence 
intervals) of only $\pm$1 day. Thus, $\Delta \tau_{gU}$ = $-$3 $\pm$ 1 days and $\Delta 
\tau_{gr}$ = 4 $\pm$ 1 days (see Table~\ref{tbl2}), indicating that $U$ is leading and $r$ is 
trailing. These band--to--band lags are consistent with delays $\tau(\lambda) \propto 
\lambda^{4/3}$ between a central driving source and standard disk rings emitting at different 
wavelengths $\lambda$ \citep[e.g.,][]{collier99}. One can easily obtain the delays $\tau$(1438 
\AA) = 6 days, $\tau$(1944 \AA) = 9 days and $\tau$(2558 \AA) = 13 days, which put the far and 
middle UV continuum sources at radii $R = c\tau/(1+z)$ of a few thousandths of a pc ($c$ is the 
speed of light). Our $gU$ and $gr$ cross--correlation studies confirm some previous evidences in 
favour of thermal reprocessing in a standard accretion disk \citep{collier01,shalyapin08}. 

For the pair $gU$, its delay distribution in Fig. 8 indicates that $U$ is leading (negative 
delay) at about the 3$\sigma$ level. Moreover, the $gr$ delay for 0957+561B is now robust because 
there are two independent estimates agreeing with each other: 4.0$^{+4.0}_{-1.7}$ days 
\citep[Apache Point Observatory data and interpolated cross--correlation function;][]{collier01}, 
and 4 $\pm$ 1 days (LRT data and $DCF$; this paper). Apart from the $DCF$ as implemented by us, 
we use a $\chi^2$ method \citep[e.g.,][]{kundic97,shalyapin08} to compare the light curves in the 
$g$ and $r$ bands. The key idea was to check the $DCF$ time lag, and to obtain effective 
parameters (flux offset and multiplicative factor; see above) for the $g$--band curve. The 
$\chi^2$ method produces a delay of 4 days, which is identical to that from the $DCF$. We also 
infer a flux offset of 0.21 mJy and a demagnification factor of 0.72. In the top panel of Fig. 9 
we display the corrected version of $g$ (squares) and the original version of $r$ (circles). 
There is very good agreement between both trends. 

The broader distributions in Fig. 8 correspond to the $XU$ (solid line), $Xg$ (dashed--dotted 
line) and $Xr$ (dotted line) pairs. Through these distributions of delays, we determine the 
shortest intervals containing 68\% of simulated values. The three 1$\sigma$ measurements are 
$\Delta \tau_{XU}$ = $-$31 $\pm$ 3 days, $\Delta \tau_{Xg}$ = $-$35.5 $\pm$ 4.5 days and $\Delta 
\tau_{Xr}$ = $-$31 $\pm$ 5 days (see Table~\ref{tbl2}). Thus, the power--law X--ray curve is 
delayed with respect to the UV--optical curves by $\sim$ 32 days. In the standard accretion 
scenario for AGNs, the observed power--law X--rays are emitted from a standard corona near the 
central black hole. It is also believed that coronal flares are thermally reprocessed into UV 
variations in the inner disk to produce the observed UV--optical variability (see Section 1). In 
this scenario we should see the X--ray fluctuations preceding the UV--optical ones, which we do 
not see. Therefore, the central driving source is not a standard corona. In the next 
subsections, we present additional evidence supporting this result, interpret the 32--day delay 
and discuss the origin of the central emission. The $\chi^2$ method for the $XU$ pair leads to a 
time lag that equal the delay estimation via the $DCF$, i.e., $-$31 days. Using 0.09 mJy and 
0.82 as flux offset and demagnification factor, respectively, we obtain a corrected version of 
$X$ (circles and line in the bottom panel of Fig. 9) that is roughly consistent with the 
original version of $U$ (triangles in the bottom panel of Fig. 9). However, some very rapid 
fluctuations in the curve $X$ do not have counterparts in the curve $U$.

\subsection{Simulations of X--ray reprocessing}
We consider a Newtonian geometrically--thin and optically--thick accretion disk, which is 
illuminated by a standard corona, i.e., an isotropic power--law X--ray source just above the 
central black hole. This lamppost model produces a radial disk temperature profile that is a 
non--linear combination of temperature resulting from viscous heating, $T_{vis}$, and that 
resulting from irradiation heating, $T_{irr}$ \citep[e.g.,][]{collier99,kazanas01}. $T_{vis}$ is 
governed by the black hole mass $M$ and the mass accretion rate $\dot{M}$, whereas $T_{irr}$ 
depends on the height $H_X$ and luminosity $L_X$ of the corona, as well as the disk reflectivity 
$A$. Here, $1-A$ represents the fraction of the X--ray radiation that is absorbed by the disk 
and reprocessed into UV--optical radiation. 

In this subsection, we focus on the lamppost model predictions for \object{0957+561}. The $z$ = 
1.41 RLAGN \object{0957+561} harbours a supermassive black hole with a mass $M = 2.5 \times 10^9 
M_{\odot}$ \citep[average of estimates through two emission lines;][]{peng06}. The corresponding 
Schwarzschild radius is given by $R_S = 2GM/c^2 = 2.5 \times 10^{-4}$ pc, where $G$ is the 
gravitational constant. For this high value of $M$, the expected $\dot{M}$ is 1--10 $M_{\odot}$ 
yr$^{-1}$ \citep{mclure04}. Moreover, $H_X < 10 R_S$, and the coronal luminosity in the 
0.01--100 keV interval can be retrieved from a suitable extrapolation of the 2--10 keV 
unabsorbed fluxes in the bottom panel of Fig. 1 and the flux--luminosity relationship in the 
presence of lens magnification. The gravitational lens magnification of the B image is 
$\mu_{lens}$ = 1.33 \citep{pelt98}, and we use a concordance cosmology \citep{spergel07}. The 
cold, neutral matter has a small reflectivity $A$ = 0.2 \citep[e.g.,][]{magdziarz95}. In Fig. 10 
we show temperature (top panel) and emission (bottom panel) profiles of the irradiated disk for 
typical values of the mass accretion rate ($\dot{M}$ = 5 $M_{\odot}$ yr$^{-1}$), as well as the 
height ($H_X = 3 R_S$) and luminosity (typical $L_X$ is based on the average unabsorbed flux; 
see above) of the corona. We find that the disk temperature is mainly due to viscous heating, so 
irradiation plays a secondary role. In addition, more than 99\% of the blackbody emission at 
$\lambda$ = 1438 \AA\ ($U$ band) and $\lambda$ = 1944 \AA\ ($g$ band) occurs within 50 $R_S$ of 
the black hole. About 97\% of the radiation at $\lambda$ = 2558 \AA\ ($r$ band) is also produced 
at $R \leq 50 R_S$. 

We note that the total luminosity of the corona over 0.01--100 keV includes photons in the 
extreme UV (EUV) and X--ray regions. However, we label this coronal emission as X--ray radiation 
for simplicity. In spite of the secondary role of the X--ray irradiation in the total heating of 
the disk in \object{0957+561}, X--ray fluctuations in the corona should cause variations in the 
disk emission \citep[e.g.,][]{collier99,kazanas01}. From the coronal luminosity associated with 
the unabsorbed variable flux in the bottom panel of Fig. 1 (see above), we can thus simulate the 
time--dependent response of the disk and the UV--optical flux variability at the observer's 
position. The key idea is to check the lamppost model by comparing simulated and observed 
records. To generate simulated light curves in the $U$, $g$ and $r$ bands, we need to know 
unabsorbed X--ray fluxes at times different to the observation epochs. Hence, we use a linear 
interpolation or a polynomial fit to produce X--ray fluxes at any epoch within the X--ray 
monitoring period, avoiding the estimation of fluxes outside this observation period. Apart from 
the emission wavelength, simulations of X--ray reprocessing also depend on seven physical 
parameters: $M$, $\dot{M}$, $H_X$, $A$, $z$, $\theta$ and $\mu$, where $\theta$ is the 
observer's latitude and $\mu$ is the AGN--observer transmission factor \citep{braga10}. At a 
given wavelength, for each physical parameter, we take either its observed value, or a 
reasonable range or no prior, and then we fit simulated fluxes to the data shown in the top 
panel of Fig. 1. 

First, X--ray fluxes during the observation period are derived from a linear interpolation, 
which allows us to produce best--fit simulated UV--optical curves (triangles in Fig. 11) after 
certain initial epochs (vertical dashed lines in Fig. 11). We assume a corona--disk radiative 
coupling with three fixed parameters: $z$ = 1.41, $A$ = 0.2 and $M = 2.5 \times 10^9 M_{\odot}$. 
The other four quantities, $\dot{M}$, $H_X$, $\theta$ and $\mu$, are treated as free parameters 
to be optimized by a $\chi^2$ technique, although we use physically motivated priors to 
accelerate the optimization process: 1 $\leq \dot{M} \leq$ 10 $M_{\odot}$ yr$^{-1}$, 1 $\leq H_X 
\leq$ 10 $R_S$ and $10\arcdeg \leq \theta \leq 80\arcdeg$. We obtain the best--fit parameter 
values $\dot{M}$ = 3 $M_{\odot}$ yr$^{-1}$, $H_X = R_S$ and $\theta = 10\arcdeg$ in the three 
$Ugr$ bands. The best fit in the $U$ band is characterized by $\mu$ = 18 and $\chi^2$/dof = 9 
(dof = 12; top panel of Fig. 11). We also derive $\mu$ = 14.5 and $\chi^2$/dof = 8 in the $g$ 
band (dof = 21; middle panel of Fig. 11), and $\mu$ = 14 and $\chi^2$/dof = 4 in the $r$ band 
(dof = 24; bottom panel of Fig. 11).  

Second, we study some variants of the analysis in the middle panel of Fig. 11 ($g$ band). 
Instead of a linear interpolation, this time we perform a polynomial fit. The best--fit curve 
(triangles in the top panel of Fig. 12) is characterized by the parameter values: $\dot{M}$ = 3 
$M_{\odot}$ yr$^{-1}$, $H_X = R_S$, $\theta = 10\arcdeg$ and $\mu$ = 14.5 ($\chi^2$/dof = 8). We 
also consider a larger reflectivity $A$ = 0.5 \citep[e.g.,][]{kazanas01}, leading to $\dot{M}$ 
= 2 $M_{\odot}$ yr$^{-1}$, $H_X = R_S$, $\theta = 10\arcdeg$ and $\mu$ = 22 ($\chi^2$/dof = 7; 
triangles in the middle panel of Fig. 12). The last variant incorporates the possibility of a 
0.1--mJy contamination generated by a source other than the disk. We then derive $\dot{M}$ = 2 
$M_{\odot}$ yr$^{-1}$, $H_X = R_S$, $\theta = 10\arcdeg$ and $\mu$ = 16 ($\chi^2$/dof = 7.5; 
triangles in the bottom panel of Fig. 12). There is no agreement between the best--fit and 
observed UV--optical light curves (see Figs. 11 and 12), which calls into question the lamppost 
model in \object{0957+561}. 

The reduced $\chi^2$ values of the best fits in Figs. 11--12, $\chi^2$/dof $>>$ 1, indicate that 
the corona--disk radiative coupling does not work in \object{0957+561}. Not even paying 
attention to the $\chi^2$, there are obvious systematic differences between the observed and 
best--fit curves, and two best--fit parameters have anomalous values. For example, $\mu = 
\mu_{lens} \mu_{dust} >$ 10 is counterintuitive because $\mu_{lens} \sim$ 1 (see above) and dust 
extinction must produce $\mu_{dust} <$ 1. However, although the corona is located on the black 
hole at $H_X = R_S$, the model does not incorporate any general relativistic effect. Thus, the 
presence of a rotating Kerr black hole may lead to disk illuminations much larger than those 
predicted by the Newtonian framework \citep{miniutti04}, avoiding the need for unphysical 
AGN--observer transmission factors. 

Additionally, the best--fit values of $\theta$ coincide with our lower limit of the observer's 
latitude range, i.e., $10\arcdeg$. This means that $\theta \leq 10\arcdeg$, in clear 
contradiction with other observations of \object{0957+561}. The observed broad emission lines 
are consistent with relatively large latitudes, so the expected dust torus around the disk does 
not intercept the line of sight to the broad--line region \citep[e.g.,][]{antonucci93}. 
Moreover, the jet morphology also suggests a relatively large latitude 
\citep[e.g.,][]{roberts85}. The small best--fit value of the latitude is an artefact related to 
the observed UV--optical and X--ray trends, and the standard model we use for connecting them. 
If the X--ray fluctuation leads the UV--optical variations, then a significant smoothing of the 
X--ray curve is required to try to reproduce the shape of the low--energy curves (see Fig. 1). 
In the lamppost model, the maximum smoothing occurs for the minimum latitude, so that very small 
(implausible) latitudes are favoured. 

\subsection{X--ray reflection}
The lamppost model involving a standard corona (see subsection 3.2) is generally supported by 
evidence of X--ray reflection by cold gas in AGNs \citep{george91}. Two reflection features are 
a fluorescent 6.4 keV Fe K$\alpha$ line and a hump in the spectrum above 10 keV. However, our 
X--ray spectra in Fig. 3 do not contain any flux excess in the 10--25 keV band (source 
rest--frame), and there is no evidence of the Fe K$\alpha$ line. This could indicate the 
presence in \object{0957+561} of a power--law X--ray source at a large height above the disk 
\citep{miniutti04}. Since the further the X--ray emission region is placed, the weaker the Fe 
K$\alpha$ line and hump will be, this interpretation naturally explains why \object{0957+561} 
does not have significant reflection features. 

The measured time lags between the X--ray record and the UV--optical curves (see 
Table~\ref{tbl2}) are also consistent with a power--law X--ray source in the direction of the 
rotation axis and located relatively far from the central black hole. Both the inner disk and 
X--ray emission region are plausibly irradiated by the driving source in the immediate vicinity 
of the black hole. Using the average X--ray/UV--optical delay of 32 days and assuming the 
observer at a typical latitude of $45\arcdeg$, the X--ray source would be at a typical height of 
$\sim$ 0.05 pc $\sim$ 200 $R_S$. This should correspond to the base of the relativistic jet 
\citep{junor99}. A schematic representation of X--ray and UV continuum sources of 
\object{0957+561} is drawn in Fig. 13. This scheme can account for all our data in the time and 
spectral domains.  

The central driving source is an elusive structure. It may be a hot torus surrounding the black 
hole \citep[e.g.,][]{rees82} and emitting EUV radiation (energies of tens of eV). This does not 
induce reflection features at keV energies. Moreover, such a source is unobservable because the 
UV spectra of both images of \object{0957+561} are totally absorbed below 2200 \AA, i.e., at $E 
>$ 13.6 eV in the source rest--frame \citep{michalitsianos93}. A second candidate is the faint 
thermal component that we detect in the Chandra X--ray spectra (see Fig. 4). This component could 
also be associated with a hot and geometrically--thick innermost accretion flow. Simulations of 
the co--evolution of supermassive black holes and their host galaxies show that the most massive 
black holes ($M > 5 \times 10^8 M_{\odot}$) are hosted by giant elliptical galaxies and are 
rapidly rotating \citep{fanidakis11}. Thus, a large fraction of the emitted very soft X--rays 
might be bent towards the disk by the strong gravitational field of a rotating Kerr black hole, 
enhancing the disk illumination and severely reducing the emission towards the observer 
\citep{miniutti04}. However, even if strong gravity effects are responsible of this anisotropic
emission, there is a serious difficulty to invoke the very soft X--rays as the central driving 
source. For example, the $g$--band optical curve in Fig. 1 has two remarkable features: (a) a 
clear decline between days 5200 and 5250, and (b) a 40\% growth from day 5250 to day 5350. These 
features are quite different from those observed in Fig. 4. Assuming a delay of $\sim$ 10 days 
between central and $g$--band variations (see subsection 3.1), the blackbody X--ray curves can 
not account for the optical variations: Fig. 4 indicates an increase in flux before day 5240, 
and constrains a possible growth over days 5240--5340 to be less than 25\% (1$\sigma$ upper limit
from linear fits to the absorbed and unabsorbed fluxes).  

\section{Conclusions}
This paper reports on a concurrent X--ray--UV--optical monitoring of the double $z$ = 1.41 RLAGN 
\object{0957+561} over six months, using the Chandra X--ray Observatory, the Swift/UVOT and the 
Liverpool Robotic Telescope. The continuum light curves for the B image of this gravitationally 
lensed AGN show prominent intrinsic fluctuations, which were predicted in advance by analysing 
optical records for the leading (A) image one year before \citep{goico09}. Our new UV--optical 
records of 0957+561B in the $U$, $g$ and $r$ bands are then used to unambiguously detect 
interband delays of several days, $U$ leading and $r$ trailing. A simple scenario can explain 
these band--to--band delays: a flaring source very close to the central supermassive black hole 
illuminates a standard accretion disk and induces UV variations at radii of $\sim$ 10--20 $R_S$. 
This finding is similar to those usually found for local AGNs at hundreds of $R_S$ from central 
driving sources \citep[e.g.,][]{collier98}.   

We also find a highly significant correlation between the brightness record for the X--ray 
emission with power--law spectrum and the three UV--optical light 
curves, with peaks corresponding to a time lag of $\sim$ 32 days, UV--optical leading. This 
correlation signal at about one month is due to the slow (tens of days in the source 
rest--frame) components of the X--ray and UV--optical light curves. The X--ray curve is thus 
significantly delayed from the UV--optical, and it is the opposite of what the accretion 
paradigm for AGNs predicts. In such a paradigm, the central source driving the variability is a 
standard corona just above the black hole. This corona emits the observed X--rays with a 
power--law spectrum, so power--law X--ray variations precede UV--optical ones. We conclude that 
the standard accretion scenario for AGNs does not account for the echo mapping in 
\object{0957+561}, since the central driving source can not be a standard corona. Detailed 
simulations of X--ray reprocessing and the lack of X--ray reflection at keV energies also rule 
out the possibility that the disk variability is driven by a standard corona. Our observations 
are consistent with power--law X--ray emission from the jet base. The emission region would 
typically be located at $\sim$ 200 $R_S$ from the central dark object (see the global scheme in 
Fig. 13). 

While the monitoring of \object{0957+561} indicates the presence of a non--standard central 
driving source, similar studies of fluctuations over tens of days in local AGNs led to a wide 
variety of results. For example, the X--ray--UV--optical variability of \object{Mrk 79} agreed 
with the accretion paradigm \citep{breedt09}. In this Seyfert nucleus, the height of the 
standard corona could vary from 10 $R_S$ (high--flux state) to 3 $R_S$ (low--flux state) over 
2.5 years. On the opposite side, a 1.5--year monitoring campaign of \object{NGC 3516} showed 
that the optical variations lead the X--rays by $\sim$ 100 days \citep{maoz00}. This result goes 
against the existence of a driving X--ray source in \object{NGC 3516}. Interestingly, the seed 
photons in \object{0957+561} could plausibly be attributed to EUV radiation produced near the 
black hole. They would drive most of the rapid variability and a part of the emission of the rest 
of the X--ray and UV continuum sources. Some of these seed photons would be thermally reprocessed 
into far and middle UV radiation in the standard disk, where viscous heating generates additional 
non--variable emission; others would be boosted to soft and hard X--rays via inverse Compton 
scattering at the base of the jet, where additional emission and fast (days) variability is 
possible (see the bottom panel of Fig. 9). Presumably, the rapid rotation of the black hole 
\citep{peng06,fanidakis11} disrupts the corona and converts it into the relativistic jet. 
However, the jet would keep a hot base as a "residual corona". 

Besides \object{0957+561}, other RLAGNs also show weak or unresolved humps and Fe K$\alpha$ 
lines \citep[e.g.,][]{grandi06}. Thus, most or all massive RLAGNs may contain power--law X--ray 
sources at relatively large heights above their central black holes, as well as central EUV 
sources driving their multiwavelength flux variations. Further continuum reverberation studies of 
low and high--redshift accretion--dominated RLAGNs are needed to check the evolution and 
universality of this non--standard accretion scenario. The combined reverberation results from 
past, current and future AGN programmes should reveal the physics in the heart of different AGNs 
over cosmic time. Finally, we note that a successful reverberation analysis requires a 
multiwavelength monitoring during a period of strong variability. At $z \geq$ 1, this task is 
much easier to plan for a gravitationally lensed AGN, since the variability of some of its images 
can be predicted in advance based on a modest optical follow--up of the lens system. Hence, our 
work opens a new window in reverberation mapping with the use of gravitational lensing. 

\acknowledgments

The authors thank several colleagues and the anonymous referee for their valuable comments on 
preliminary versions of this paper. We also thank the staff of the Chandra X-ray Observatory 
(CXO), Liverpool Robotic Telescope (LRT) and Swift Multi-wavelength Observatory (SMO) for their 
kind interaction and support during the preparation of the project and the observation period. 
The CXO Center is operated by the Smithsonian Astrophysical Observatory for and on behalf of the 
National Aeronautics Space Administration (NASA) under contract NAS803060. The LRT is operated on 
the island of La Palma by Liverpool John Moores University in the Spanish Observatorio del Roque 
de los Muchachos of the Instituto de Astrof\'{\i}sica de Canarias (ORM-IAC) with support from the 
UK Science and Technology Facilities Council. The SMO is supported at Penn State University by 
NASA contract NAS5-00136. This article is also based on observations made with the Nordic Optical 
Telescope, which is operated on the island of La Palma jointly by Denmark, Finland, Iceland, 
Norway, and Sweden, in the Spanish ORM-IAC. This research has been supported by the Spanish 
Department of Science and Innovation grants AYA2007-67342-C03-02 and ESP2006-13608-C02-01, and 
University of Cantabria funds. 


\clearpage

\begin{figure}
\epsscale{0.7} 
\plotone{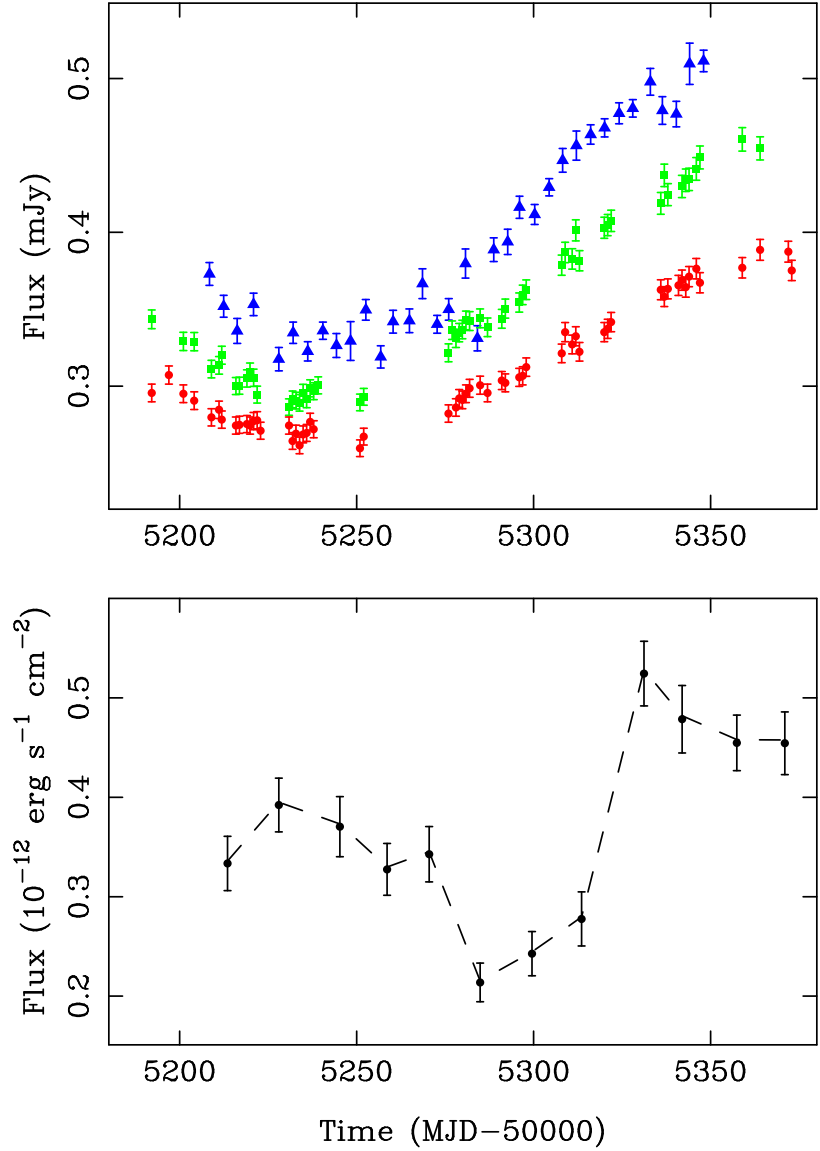}
\caption{Top: Optical and UV light curves of 0957+561B from late 2009 to mid 2010. Triangles 
(UVOT, $U$ band), squares (LRT, $g$ band) and circles (LRT, $r$ band) are observed fluxes in mJy, 
associated with sources emitting at $\lambda$ = 1438 \AA\ (8.6 eV), 1944 \AA\ (6.4 eV) and 2558 
\AA\ (4.8 eV), respectively. The optical light curves are vertically shifted by $-$0.13 mJy ($g$ 
band) and $-$0.25 mJy ($r$ band) to improve visual comparison. Bottom: Chandra X--ray light 
curve of 0957+561B from early to mid 2010. Circles are observed fluxes of the power--law 
X--ray source (range 2--10 keV in the source rest--frame), while the dashed line traces the time 
evolution of the unabsorbed flux in the same energy range.}
\end{figure}

\clearpage

\begin{figure}
\plotone{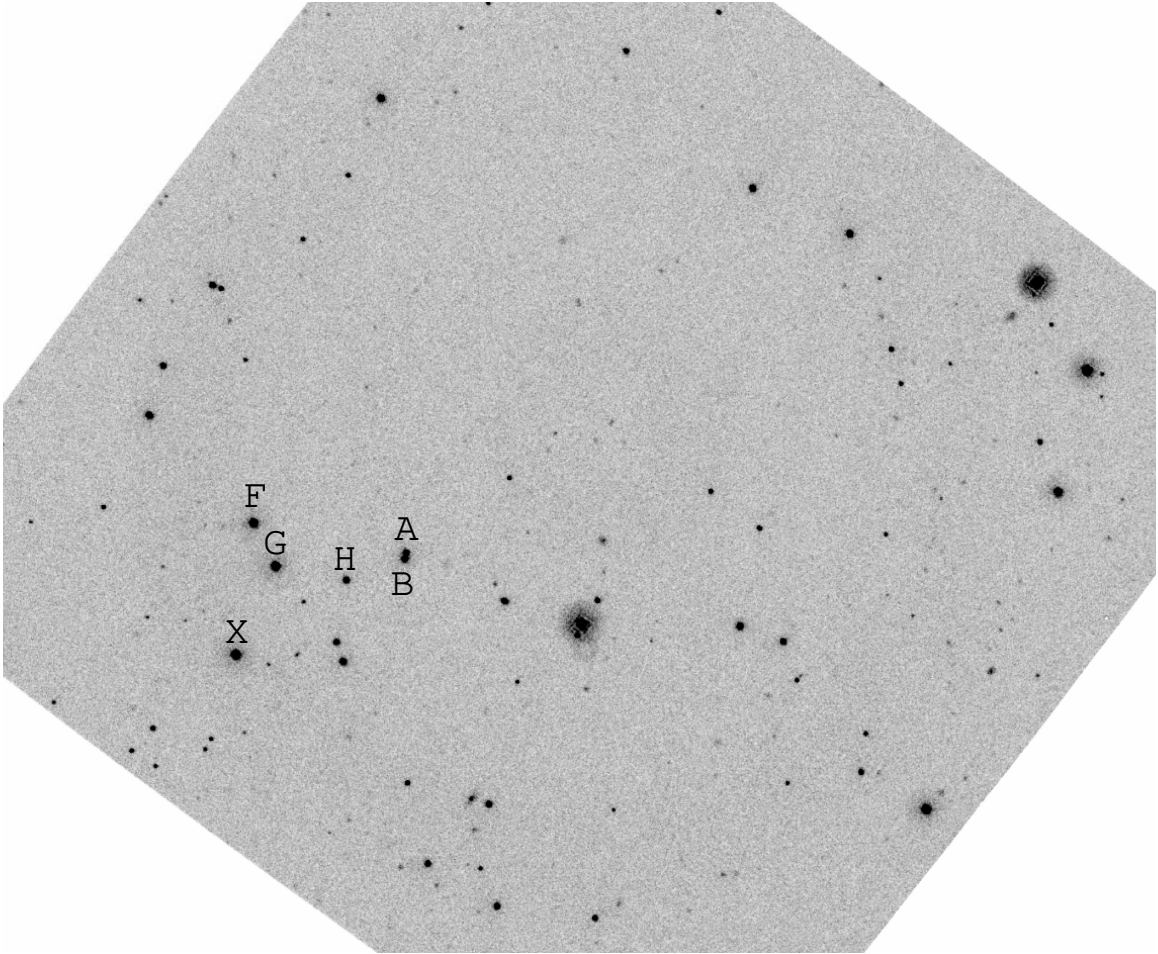}
\caption{Near UV Swift/UVOT frame of the field of 0957+561. This 1092 s exposure was taken on 
2010 May 12. The AGN images are labelled A and B, whereas some bright stars close to the AGN are 
labelled H, X, G and F \citep[e.g., see Fig. 1 of][]{ovaldsen03}. There are many additional 
bright sources within the $16\arcmin \times 16\arcmin$ field of view, so the 5--month monitoring 
campaign can be used to detect new UV variables or to characterize the short--term variability 
of known variables.} 
\end{figure}

\clearpage

\begin{figure}
\epsscale{1.0} 
\plotone{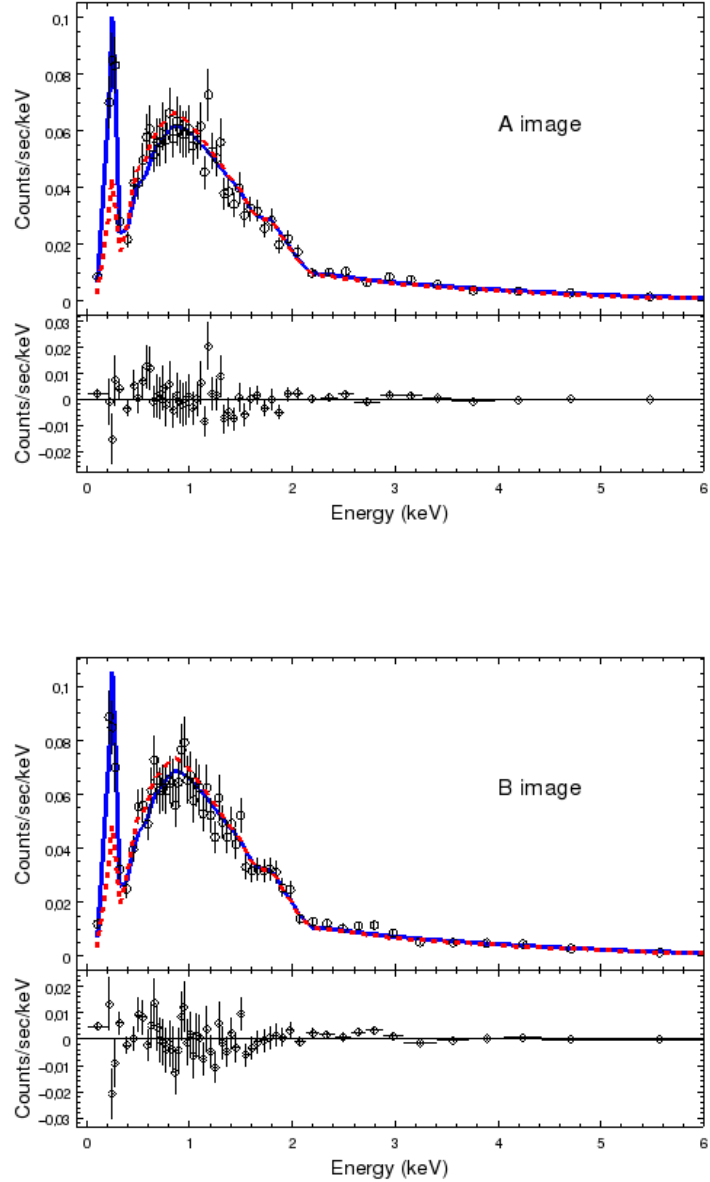}
\caption{Chandra X--ray spectra of 0957+561A and 0957+561B. These spectra are stacked from all
observing epochs. An absorbed power--law model produces poor fits to the data ($\Gamma$ = 1.9 
and $\chi^2$/dof = 2.8; dashed lines), whereas a power--law plus blackbody and Galactic 
absorption model notably improves the fits ($\Gamma$ = 1.78, $kT$ = 0.08 keV and $\chi^2$/dof 
$\sim$ 1; solid lines). We introduce a blackbody emission to explain the soft excess at energies 
$<$ 0.5 keV (the horizontal axis represents energy in the observer rest--frame). The residuals 
for each quasar image (data –- solid line) are displayed in the bottom subpanels. We do not find 
evidence for a Fe K$\alpha$ line at $\sim$ 2.7 keV (6.4 keV in the source rest--frame) and a hump 
above 4 keV ($>$ 10 keV in the source rest--frame).} 
\end{figure}

\clearpage

\begin{figure}
\epsscale{0.7} 
\plotone{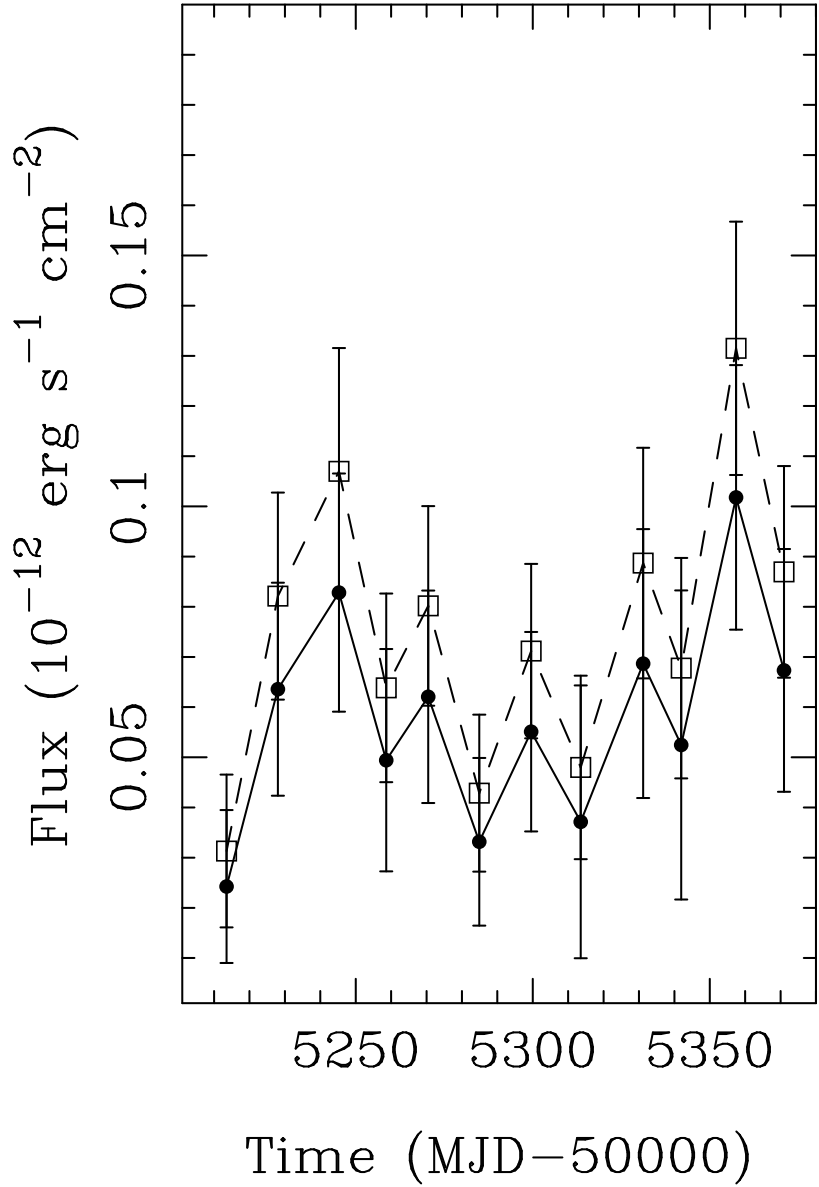}
\caption{Blackbody X--ray light curves of 0957+561B. Circles and squares are observed (absorbed) 
and unabsorbed fluxes at energies below 1 keV in the source rest--frame. The blackbody component 
with $kT$ = 0.08 keV is much fainter than the power--law component (see the bottom panel of Fig. 
1). The data points have large relative errors that significantly exceed those in the light curves 
in Fig. 1.} 
\end{figure}

\clearpage

\begin{figure} 
\plotone{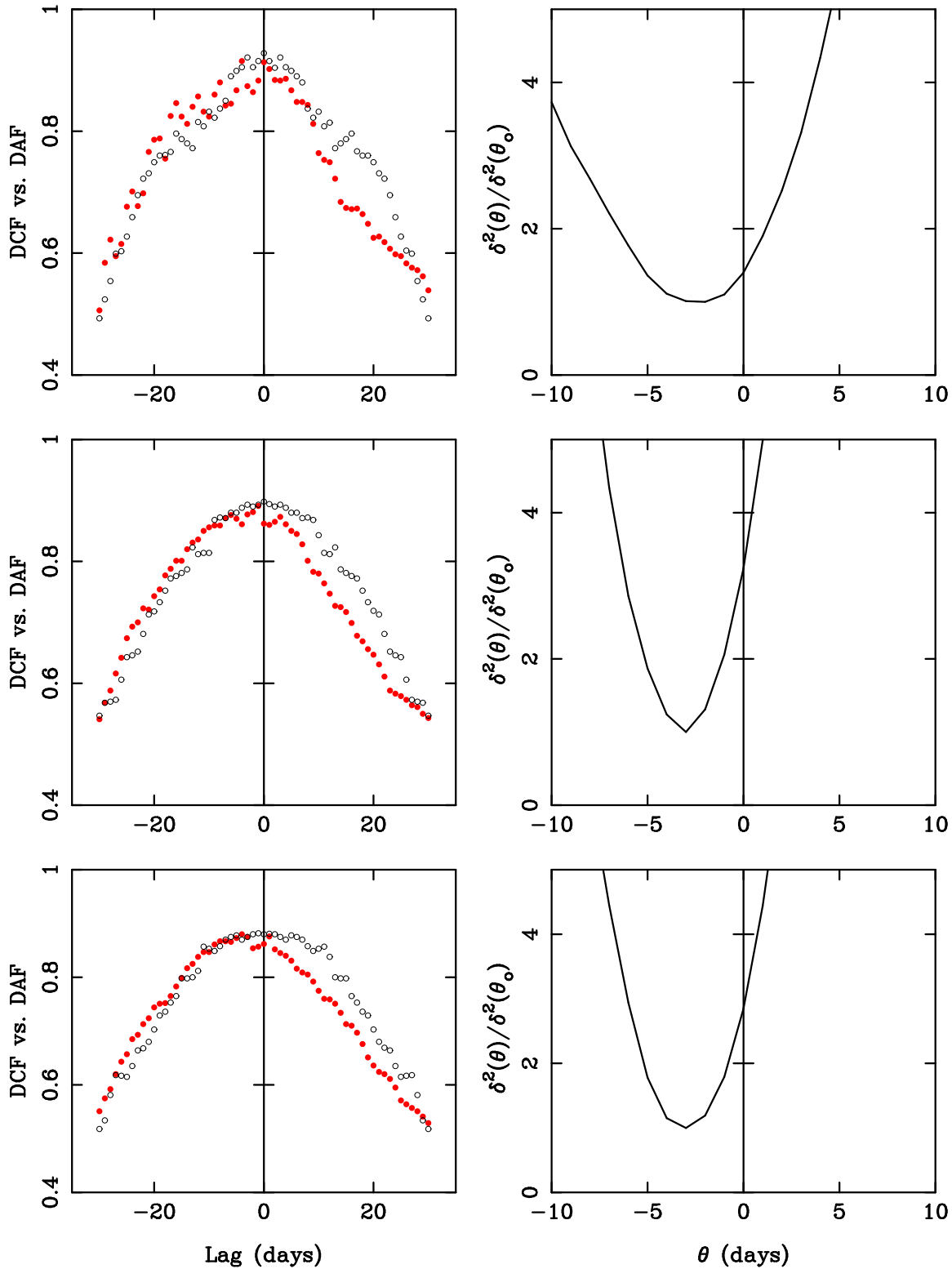}
\caption{Time delay between the $g$ and $U$ bands in 0957+561. The left panels show the 
comparisons between the $DCF$ (filled circles) and the $DAF$ (open circles) for three values of 
the bin semisize $\alpha$. While the $DCF$ is the $gU$ cross--correlation, the $DAF$ is the 
average of the $gg$ and $UU$ autocorrelations. In each left panel, there is clear evidence of a 
$gU$ time delay, since the $DAF$ peak should be shifted to the left by several days to optimally 
match the $DCF$ delay--peak. Possible values of this time shift $\theta$ versus the associated 
$\delta^2$ values normalised by its minimum value $\delta^2(\theta_0)$ are displayed in the 
corresponding right panel (see main text). Top: $\alpha$ = 7 days. Middle: $\alpha$ = 10 days.
Bottom: $\alpha$ = 12 days.}
\end{figure}

\clearpage

\begin{figure} 
\plotone{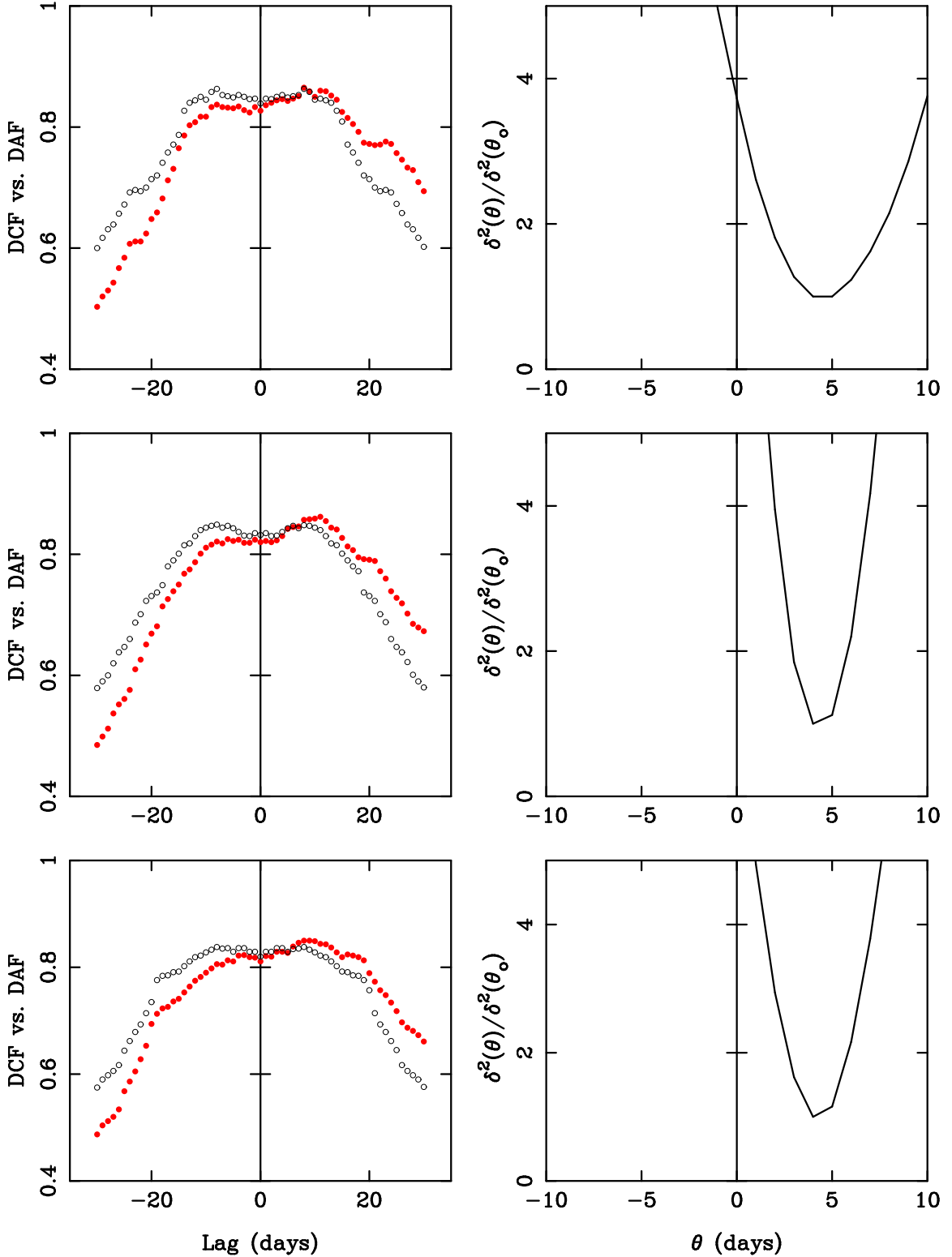}
\caption{Time delay between the $g$ and $r$ bands in 0957+561. The $DCF$ (filled circles) is the 
$gr$ cross--correlation, the $DAF$ (open circles) is the average of the $gg$ and $rr$ 
autocorrelations, and $\delta^2(\theta)$ is the function to be minimized (see Fig. 5 and main 
text). In the top, middle and bottom panels, we can observe the presence of a time shift between 
the $DCF$ and $DAF$, so the $DAF$ should be shifted to the right by a few days to optimally match 
the $DCF$ peak. Top: $\alpha$ = 15 days. Middle: $\alpha$ = 18 days. Bottom: $\alpha$ = 20 days.}
\end{figure}

\clearpage

\begin{figure} 
\plotone{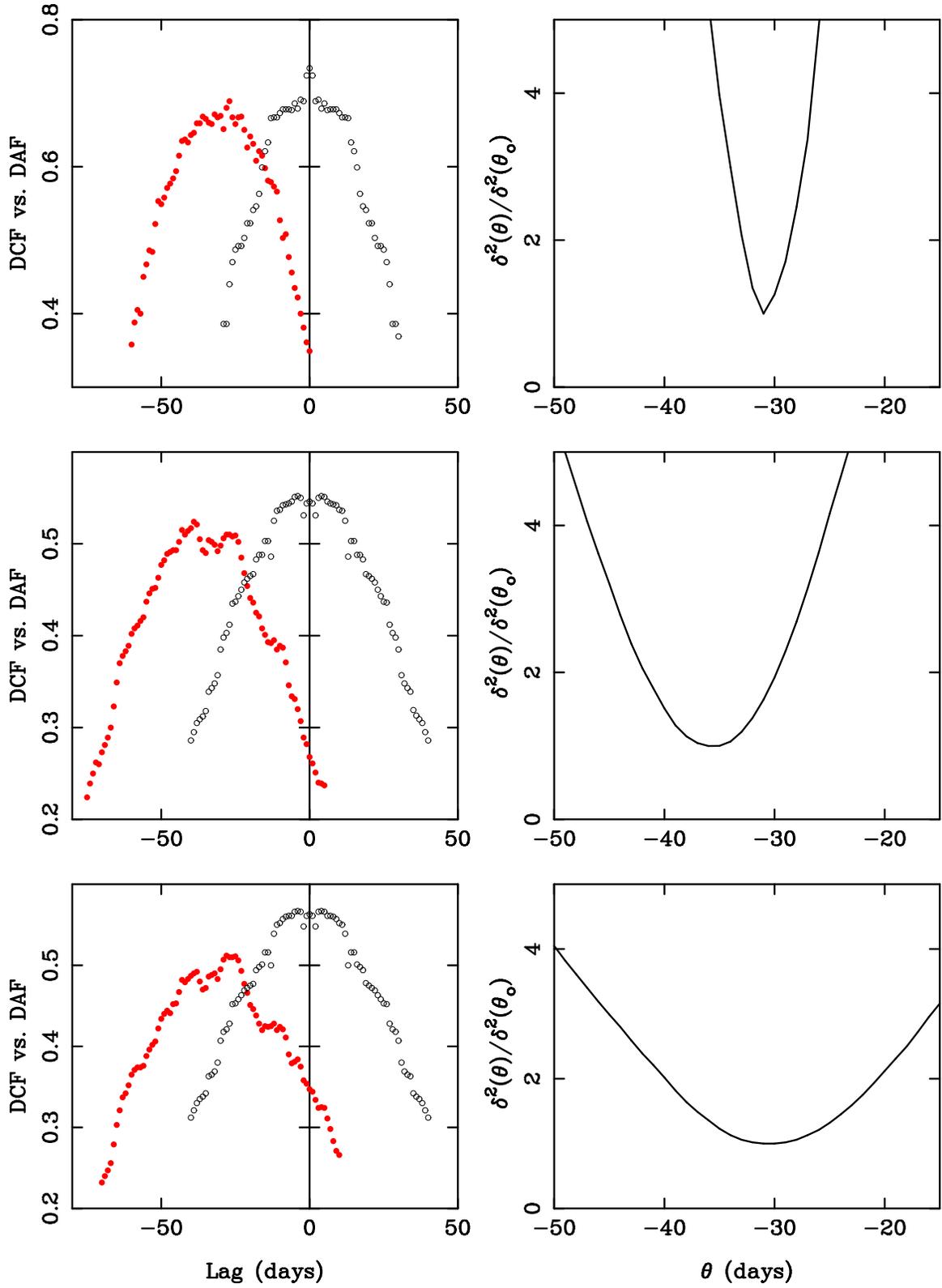}
\caption{Time delay between the power--law X--ray and UV--optical records in 0957+561 (see Figs. 
5--6 and main text for basic notation and meaning of symbols). In order to optimally match the 
$DCF$ peak, the $DAF$ should be shifted to the left by about one month. Top: $XU$ comparison with 
$\alpha$ = 30 days. Middle: $Xg$ comparison with $\alpha$ = 45 days. Bottom: $Xr$ comparison with 
$\alpha$ = 45 days.}
\end{figure}
 
\clearpage

\begin{figure}
\epsscale{0.7} 
\plotone{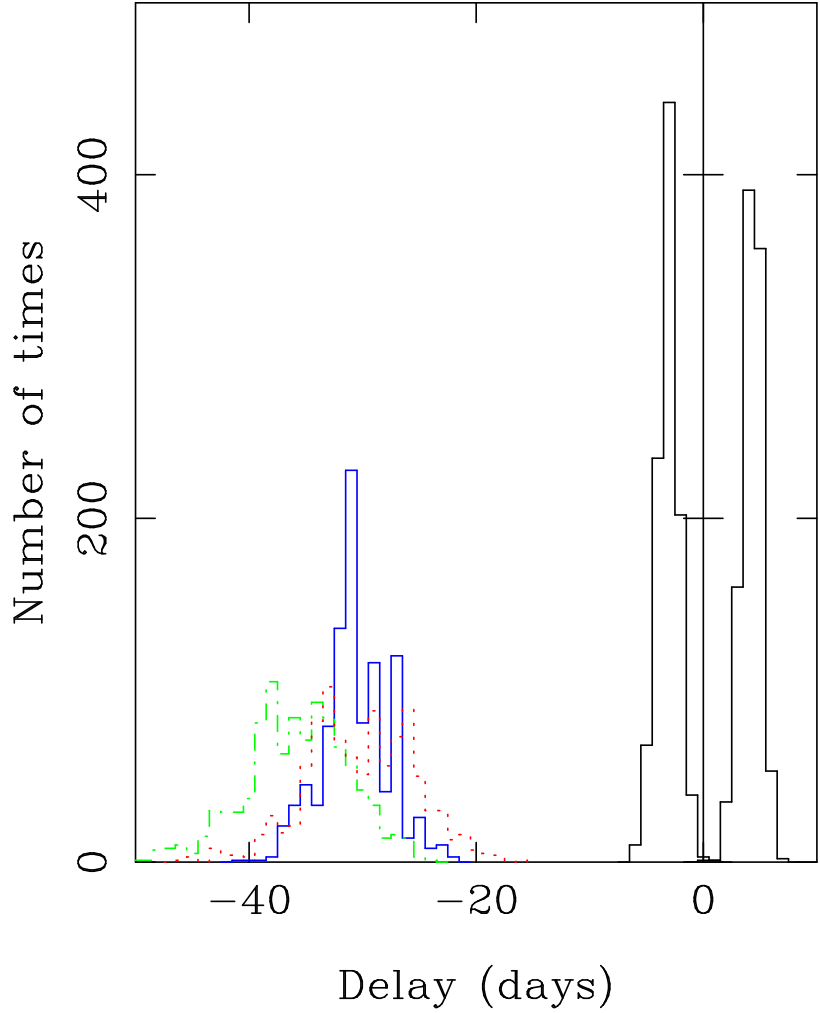}
\caption{Delay distributions. From 1000 simulated repetitions of the original experiment with the 
LRT, Swift/UVOT and Chandra, we obtain 1000 delay values for each pair of curves $gU$ (narrow 
histogram around a delay of $-$3 days), $gr$ (narrow histogram around a delay of 4 days), $XU$ 
(broad histogram drawn with a solid line), $Xg$ (broad histogram traced by a dashed-dotted line) 
and $Xr$ (broad histogram traced by a dotted line). These distributions allow us to determine time 
delay errors.} 
\end{figure}
 
\clearpage

\begin{figure} 
\epsscale{0.65} 
\plotone{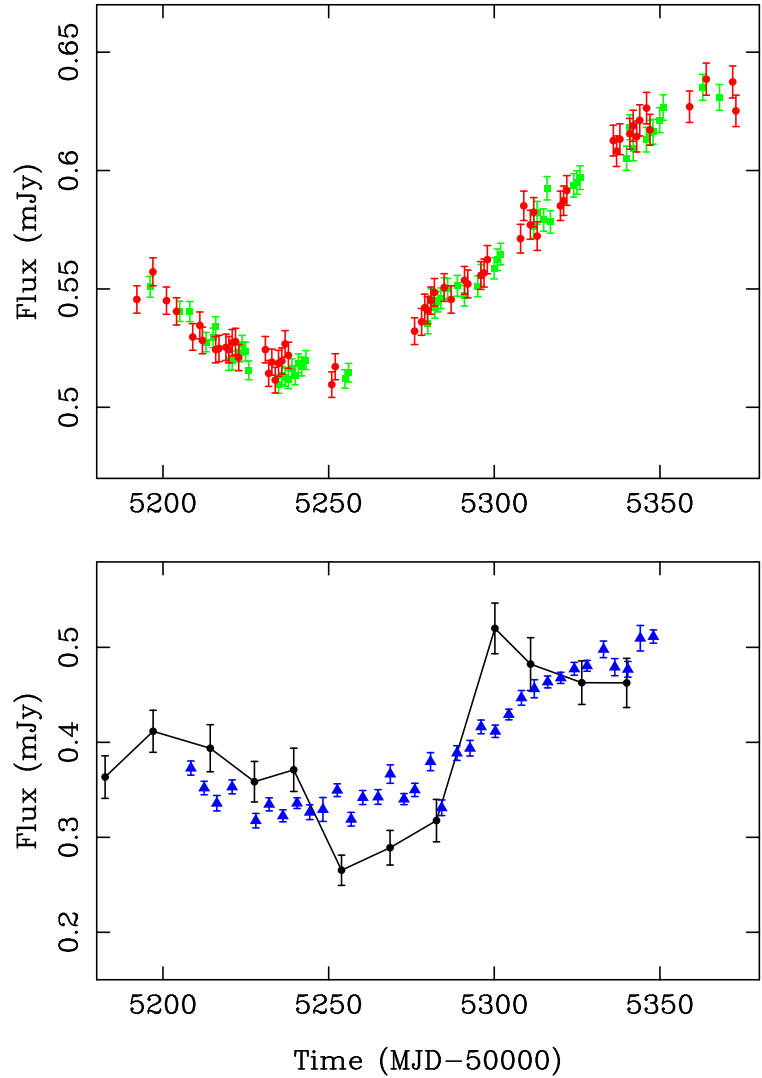}
\caption{$\chi^2$ comparisons. We use a $\chi^2$ method to compare two light curves $a$ and $b$, 
obtaining a time delay, a flux offset and a multiplicative factor for the curve $a$. Once these 
three parameters are known, we put together the shifted and (de)magnified version of the curve 
$a$, and the curve $b$. Top: $gr$ light curves. We infer a delay of 4 days, a flux offset of 0.21 
mJy and a demagnification factor of 0.72. The corrected version of $g$ and the original version 
of $r$ are represented by squares and circles, respectively. Bottom: $XU$ light curves. The 
$\chi^2$ technique leads to a delay of $-$31 days. This time lag, a flux offset of 0.09 mJy and a 
demagnification factor of 0.82 are used to make the corrected version of $X$ (circles and line). 
The original version of $U$ is traced by triangles.}
\end{figure}

\clearpage

\begin{figure} 
\epsscale{0.7} 
\plotone{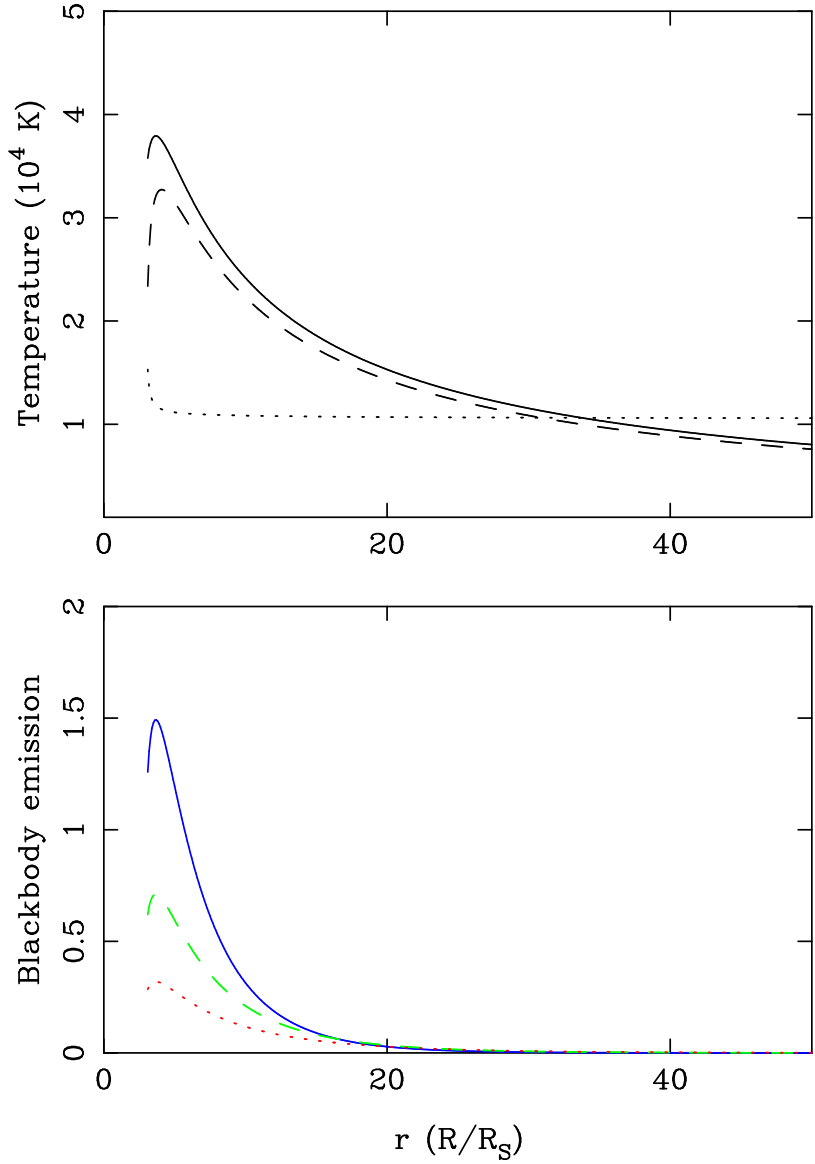}
\caption{Top: Radial temperature profile for an irradiated accretion disk in 0957+561. The dashed 
and solid lines represent $T_{vis}$ and the total temperature $T = (T_{vis}^4 + 
T_{irr}^4)^{1/4}$, respectively. The dotted line describes the ratio $T/T_{vis}$. As usual, the 
inner radius of the disk is set to 3 $R_S$. Bottom: Blackbody emission profile of the disk at 
different wavelengths. The solid, dashed and dotted lines correspond to blackbody sources at $\lambda$ 
= 1438 \AA, 1944 \AA\ and 2558 \AA, respectively. The radial emission profile (Planck function 
in arbitrary units) of the three sources is enhanced in the vicinity of the innermost ring at $R 
= 3 R_S$.}
\end{figure}

\clearpage

\begin{figure}
\epsscale{0.6} 
\plotone{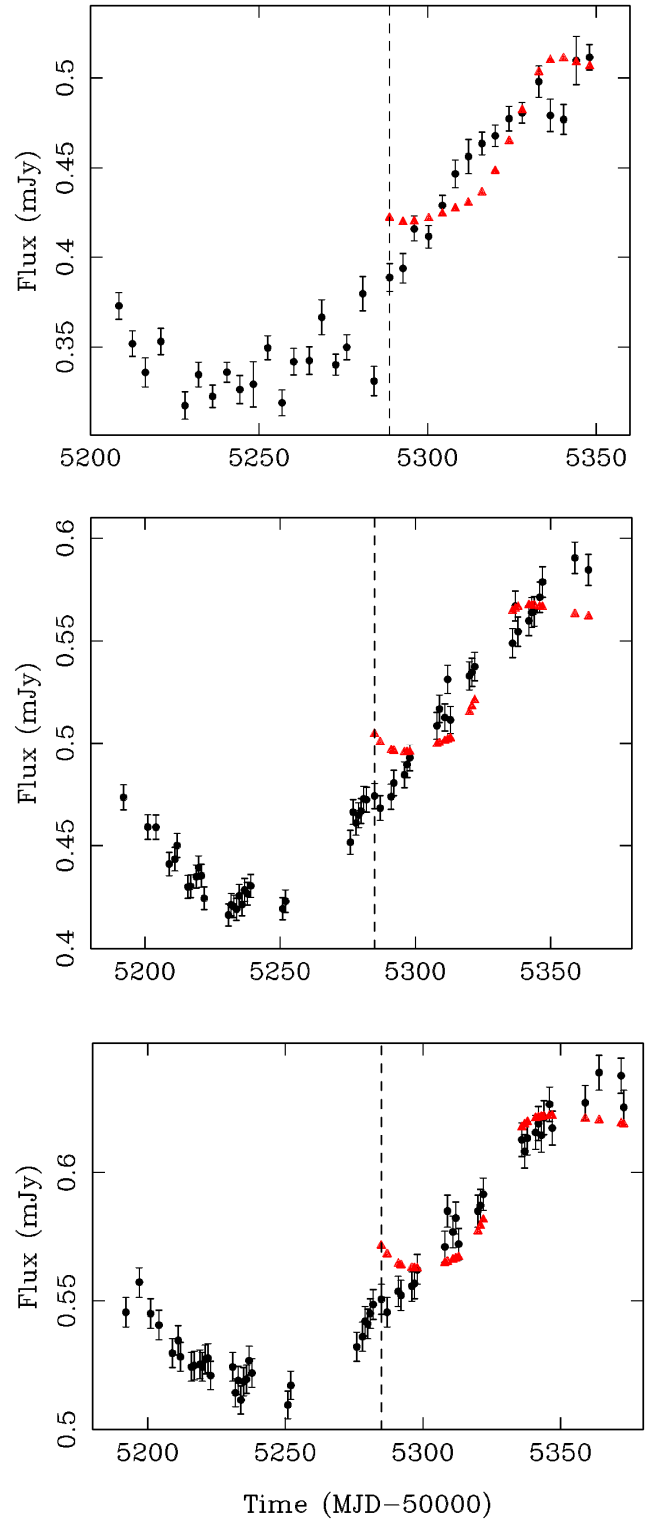}
\caption{X--ray reprocessing in 0957+561. X--ray fluxes at any epoch within the X--ray monitoring 
period are derived from a linear interpolation procedure. These fluxes are then used to create 
simulated UV--optical records, and to carry out $\chi^2$ fits between the simulated and observed 
curves (see main text). The observed $Ugr$ fluxes are represented by circles, and the best--fit 
simulated $Ugr$ curves are traced by the triangles after the vertical dashed lines. Top: $U$ 
band. Middle: $g$ band. Bottom: $r$ band.}
\end{figure}

\clearpage

\begin{figure} 
\epsscale{0.6} 
\plotone{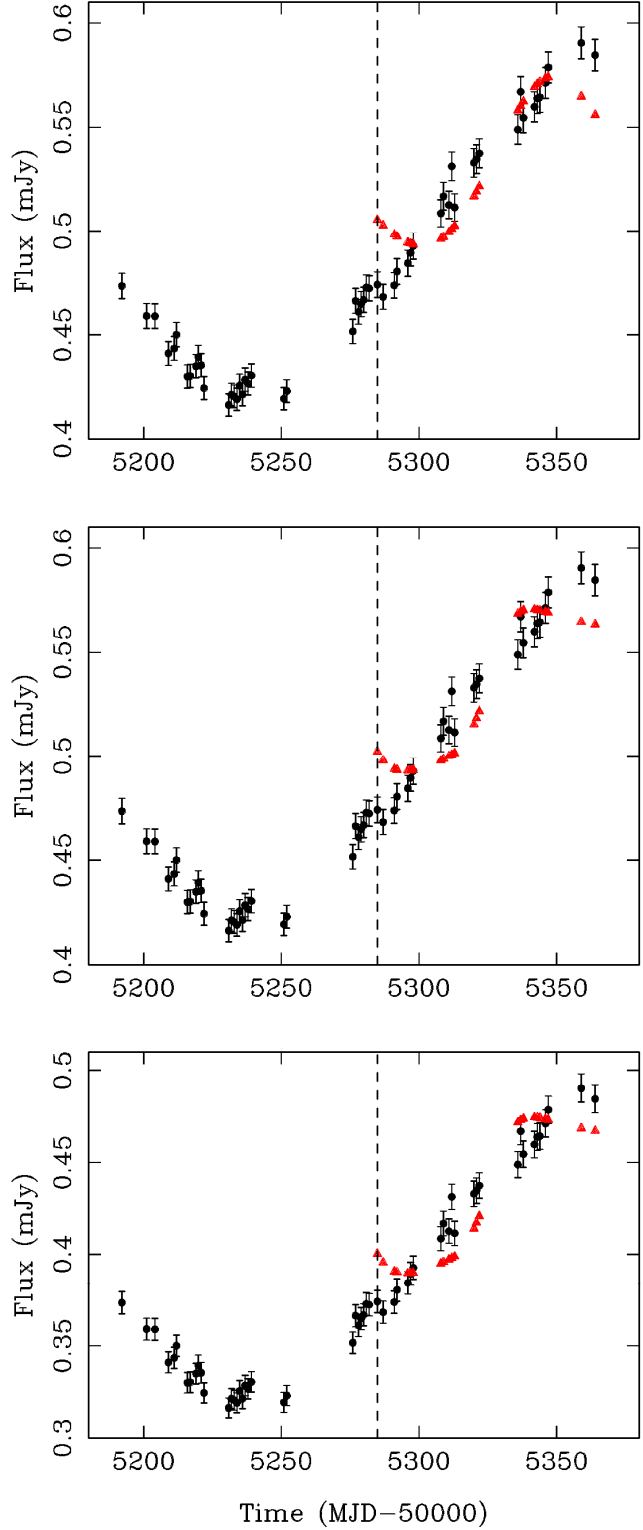}
\caption{Thermal reprocessing of X--rays into $\lambda$ = 1944 \AA\ ($g$ band) radiation in 
0957+561. We show some variants of the analysis in the middle panel of Fig. 11. The observed 
$g$--band record is traced by circles, while triangles describe the best--fit curves. Top: X--ray 
fluxes during the observation period are estimated from a 7--degree polynomial resulting from a 
good fit to the X--ray data. Middle: The disk reflectivity or albedo is increased to $A$ = 0.5. 
Bottom: We subtract a possible constant contamination from the observed $g$--band curve. The 
constant extra--disk flux is set to 0.1 mJy.}
\end{figure}

\clearpage

\begin{figure}
\epsscale{0.9} 
\plotone{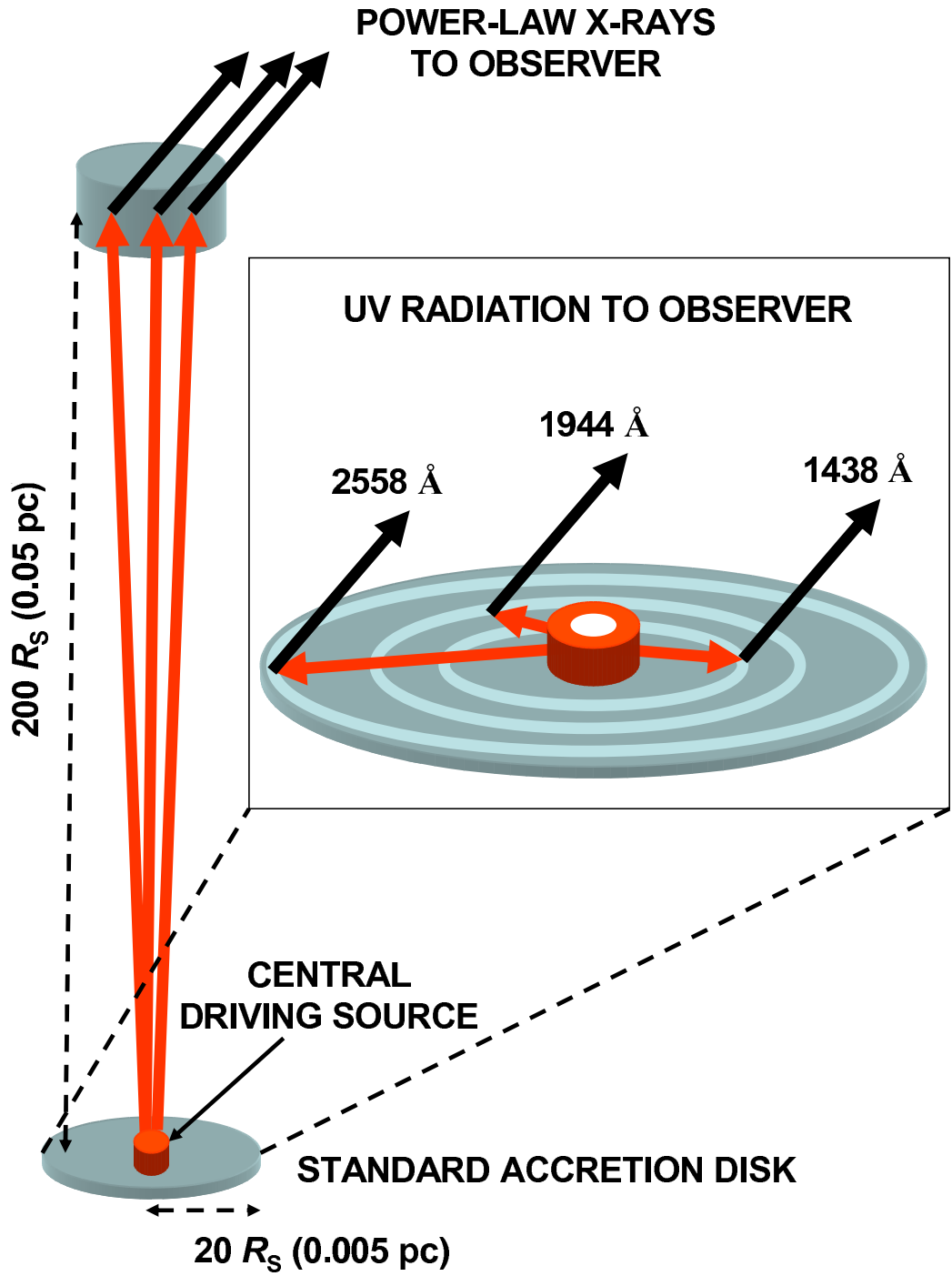}
\caption{Schematic interpretation of the results (see main text).}
\end{figure}

\clearpage

\begin{deluxetable}{cccc}
\tablecaption{0957+561 monitoring data.\label{tbl1}}
\tablewidth{0pt}
\tablehead{
\colhead{Instrumentation} & \colhead{Obs. mode\tablenotemark{a}} & \colhead{Start/end dates} & 
\colhead{Exposures (ks)\tablenotemark{b}} 
}
\startdata
LRT/RATCam/$gr$ filters\tablenotemark{c}    & IMAG--M & 2009 Dec 26/2010 Jun 25    & 0.12 \\           
Swift/UVOT/CCD/$U$ filter\tablenotemark{d}  & IMAG--M & 2010 Jan 12/2010 Jun 1     & $\sim$ 0.5 \\  
Chandra/HRMA/ACIS--S3\tablenotemark{e}      & SPEC--M & 2010 Jan 17/2010 Jun 23    & $\sim$ 3 \\ 
\enddata
\tablenotetext{a}{The basic observing modes were imaging (IMAG) and spectroscopy (SPEC). The label 
--M means that the observation scheme was repeated on a regular basis between the start and end 
dates, i.e., monitoring campaign.}
\tablenotetext{b}{The exposure times in ks refer to each filter or spectrometer on a single night.}
\tablenotetext{c}{Programmes CL09B03 and CL10A02.}
\tablenotetext{d}{Target ID 31567.}
\tablenotetext{e}{Programme 10708333.}
\end{deluxetable}


\begin{deluxetable}{cccc}
\tablecaption{Time lags (in days) between multiwavelength records of 0957+561.\label{tbl2}}
\tablewidth{0pt}
\tablehead{
\colhead{$a$} & \colhead{$b$} & \colhead{$\Delta \tau_{ab}$} & \colhead{Comment} 
}
\startdata
$g$  	&	$U$ 	&	$-$3 $\pm$ 1	&	$U$ leading\\           
	&  	$r$  	&	4 $\pm$ 1		&	$g$ leading\\
$X$  	&	$U$ 	&	$-$31 $\pm$ 3 	&	$U$ leading\\           
	&  	$g$  	&	$-$35.5 $\pm$ 4.5 &	$g$ leading\\
  	&	$r$ 	&	$-$31 $\pm$ 5 	&	$r$ leading\\           
\enddata
\tablecomments{We compare records in the $Ugr$ UV--optical bands and the X--ray region. 
$\Delta \tau_{ab} = \tau_b - \tau_a$, and all measurements are 68\% confidence intervals.}
\end{deluxetable}

\end{document}